\documentclass[aps,prb,twocolumn,showpacs,amsmath,superscriptaddress,floatfix]{revtex4-1}

\usepackage{amsmath}
\usepackage{epsfig}
\usepackage{graphicx}
\usepackage{graphics}
\usepackage{url}
\usepackage[polish,english]{babel}
\usepackage{dcolumn}
\usepackage{color}

\newcommand{\be}{\begin{equation}}
\newcommand{\ee}{\end{equation}}
\newcommand{\ba}{\begin{eqnarray}}
\newcommand{\ea}{\end{eqnarray}}
\renewcommand{\vec}[1]{\textbf{#1}}
\newcommand{\ka}{{\vec{k}}}

\newcommand{\qu}{{\vec{q}}}

\newcommand{\dt}{{\mathrm{d}t}}
\renewcommand{\dag}{^{\dagger}}
\newcommand{\ndag}{^{\phantom{\dagger}}}
\newcommand{\bra}[1]{\left\langle#1\right|}                 
\newcommand{\ket}[1]{\left|#1\right\rangle}                 
\newcommand{\bks}[1]{\langle #1 \rangle}                    
\newcommand{\fref}[1]{(\ref{#1})}
\newcommand{\X}[2]{{e$_{#1}$h$_{#2}$}}
\newcommand{\ls}{{\text{l}s}}
\newcommand{\lp}{{\text{l}p}}
\newcommand{\ld}{{\text{l}d}}
\newcommand{\lf}{{\text{l}f}}
\newcommand{\us}{{\text{u}s}}
\newcommand{\up}{{\text{u}p}}
\newcommand{\ud}{{\text{u}d}}

\newcommand{\Tag}[1]{{\normalsize{\textcircled{\textsf{{\scriptsize{#1}}}}}}}

\begin{document}

\title{Excitons in quantum dot molecules: Coulomb coupling, spin-orbit effects and
phonon-induced line broadening}

\author{J.~M.~Daniels}
\affiliation{Institut f\"ur Festk\"orpertheorie, Westf\"alische Wilhelms-Universit\"at
M\"unster, Wilhelm-Klemm-Str.~10, 48149 M\"unster, Germany}

\author{P.~Machnikowski}
\affiliation{Institute of Physics, Wroc\lpb{}aw University of Technology, 50-370
Wroc\lpb{}aw, Poland}

\author{T.~Kuhn}
\affiliation{Institut f\"ur Festk\"orpertheorie, Westf\"alische Wilhelms-Universit\"at
M\"unster, Wilhelm-Klemm-Str.~10, 48149 M\"unster, Germany}

\date{\today}

\begin{abstract}
    Excitonic states and the line shape of optical transitions in coupled quantum
    dots (quantum dot molecules) are studied theoretically. For a pair of
    electrically tunable, vertically aligned quantum dots we investigate the
    coupling between spatially direct and indirect excitons caused by different
    mechanisms such as tunnel coupling, Coulomb coupling, coupling due to the
    spin-orbit interaction and coupling induced by a structural asymmetry. The
    peculiarities of the different types of couplings are reflected in the
    appearance of either crossings or avoided crossings between direct and
    indirect excitons, the latter ones being directly visible in the absorption
    spectrum. We analyze the influence of the phonon environment on the spectrum
    by calculating the line shape of the various optical transitions including
    contributions due to both pure dephasing and phonon-induced transitions
    between different exciton states. The line width enhancement due to
    phonon-induced transitions is particularly pronounced in the region of an
    anticrossing and it strongly depends on the energy splitting between the two
    exciton branches.
\end{abstract}

\pacs{73.21.La, 78.67.Hc, 63.20.kd}

\maketitle

\section{Introduction}
Structures consisting of two closely spaced quantum dots (QDs) have attracted
much interest in the past years both in
experiment\cite{Bayer2001,Krenner2005,Stinaff2006,Bracker2006,Robledo2008,Goryca2009,
Shulmman2012} and in
theory.\cite{Bester2004,Schliwa2001,Szafran2005,Szafran2008,Gawarecki2010}
Carriers in these nanostructure are not strictly confined to one dot but may
be delocalized over both dots, which is often described in terms of a tunnel
coupling between the individual
dots.\cite{Bryant1993,Bracker2006,Scheibner2008} This coupling causes the
formation of bonding and anti-bonding states in the double dot systems, hence
they are often referred to as quantum dot molecules (QDMs). External fields
can be used to effectively manipulate carrier
localizations\cite{Krenner2005,Ortner2005PRL} and to characterize different
exciton states,\cite{Ortner2003,Lyanda-Geller2004,Ortner2005PRB,
Mueller2011,Heldmaier2012} which is essential for many applications.

Since QDs are embedded in a solid state matrix, the crystal environment plays
a crucial role in the carrier kinetics in a QDM. Phonons provide a source of
dephasing in single QDs\cite{Takagahara1999,Krummheuer2002} as well as in
coupled QDs.\cite{Borri2003,Stavrou2005} They give rise to the relaxation of
the exciton, although this is restricted by the phonon bottleneck
effect.\cite{Benisty1995} But even if transition rates are strongly reduced
because of the lack of final states in the energy region where the coupling
to phonons is efficient, phonon-induced pure dephasing has a pronounced
influence on the line shape of absorption or emission
spectra.\cite{Besombes2001,Krummheuer2002} For a QDM also phonon-mediated
relaxation between bonding and anti-bonding states as well as phonon-assisted
tunneling or interdot relaxation may take
place.\cite{Lopez-Richard2005,Vorojtsov2005,
Wu2005,Grodecka2008,Gawarecki2010,Gawarecki2012}

Experimental studies of exciton spectra in QDMs placed in an axial electric
field show characteristic features that are qualitatively different from the
quantum confined Stark shift behavior observed in individual QDs. The most
pronounced features emerging in the QDM spectra are resonances
(anticrossings) that appear at certain intersections between spatially direct
and indirect exciton states (that is, states with the electron and the hole
residing in the same QD or in different QDs,
respectively).\cite{Scheibner2008,Mueller2011} Recent experiments
investigated also the phonon-assisted interdot tunneling between the direct
and indirect configurations.\cite{Mueller2012} Additionally, it was shown,
that the phonon-induced relaxation time of an exciton at a tunnel-resonance
(anticrossing) exhibits an oscillatory behavior due to an interplay between
the phonon wavelength and the distance between the
dots.\cite{Wijesundara2011,Rolon2012} Theoretical modeling of exciton states
in QDMs reproduces the observed spectral features, including the structure of
the lowest-energy states \cite{Szafran2001} and their behavior in an axial
electric field,\cite{Sheng2002,Janssens2002} as well as the tunnel-related
anticrossings between the spatially direct and indirect configurations as a
function of the electric field, where the crucial role is played by the
electron-hole Coulomb interaction.\cite{Szafran2005,Szafran2007,Szafran2008}
Couplings induced by the spin-orbit coupling (SOC) have been studied in
laterally coupled quantum dots.\cite{Stano2005} Several theoretical studies
have also addressed phonon-related processes in QDMs, focussing on single-
\cite{Lopez-Richard2005,Vorojtsov2005, Wu2005,Gawarecki2010} and two-electron
\cite{Gawarecki2010,Grodecka2008} as well as single-hole systems
\cite{Gawarecki2012} and excitons.\cite{Muljarov2005}

In this paper, we present a systematic study of excitons in QDM structures,
which are tunable by an external electric field.  We extend previous research
by investigating avoided exciton level-crossing mediated by Coulomb coupling
with higher exciton states. We also consider couplings between exciton states
with different values of the angular momentum which can be induced either by
the SOC or by a breaking of the cylindrical symmetry of the confinement
potential. To calculate the line shape of the transitions in the absorption
spectrum we take into account the coupling to acoustic phonons including both
the phenomena of pure dephasing as well as phonon-induced transitions to
other exciton states.

From our calculations we find that Coulomb interaction is crucial for
understanding the sequence of exciton states and tunnel resonances. We show
that besides modifying the sequence of the states the Coulomb interaction
also leads to the appearance of additional anticrossings between two
configurations in which not only one of the carriers (electron or hole)
resides in different dots but also the other carrier is in different states
within one dot. Such two-particle states are not connected by the usual
single-particle tunneling and, therefore, would not give rise to a resonant
anticrossing in the absence of the off-diagonal part of the Coulomb coupling.
The effective coupling between the two states forming such a resonance is
found to be of comparable magnitude as the usual tunnel coupling. For the
combined system of exciton and phonons, we investigate phonon-assisted
tunneling and the line shape of absorption as a function of temperature. In
addition, we study the effect of the SO coupling in the area of level
crossings (corresponding to the degeneracy points between states of different
angular momentum). We show that it leads to a qualitative change of the
system spectrum by opening a resonant anticrossing between spin-bright and
spin-dark states with different angular momenta. A comparable effect can be
caused by a misalignment of the dots or by deviations from a rotational
symmetry of the QDs, which breaks the conservation of angular momentum.

The paper is organized as follows: In Sec.~\ref{Theo} we summarize the
underlying theory of the calculations. We start with the model for the
carrier wave-functions and the exciton states. Then, the coupling of the
exciton to the phonon system is described and a Green's function formalism
for the calculation of absorption spectra is introduced, followed by the
introduction of spin-orbit coupling. In the same way the results are
presented in Sec.~\ref{Res}. We start with the exciton energy-level structure
and a discussion of the absorption spectra of a typical QDM. Then, different
types of crossing and avoided crossings are discussed as well as the impact
of the phonon environment. Finally the coupling between excitons states with
different angular momentum either due to the SOC or due to a broken
cylindrical symmetry of the confinement potential is discussed. The paper
concludes with a brief summary and some final remarks in Sec.~\ref{Conc}.

\section{Model} \label{Theo}
\subsection{Exciton States} \label{Theo:ExState}
\begin{figure}[t]
    \includegraphics[width=0.8\columnwidth]{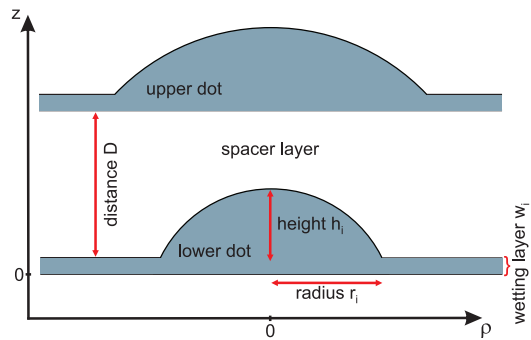}
    \caption{ \label{fig:scheme_molecule} (Color online) Sketch of the quantum dot molecule.}
\end{figure}
We consider a QDM formed by two vertically stacked QDs of a lens-like shape
in a GaAs matrix. The QDs are modeled as spherical segments of
In$_x$Ga$_{1-x}$As with homogenous indium fraction $x$, base radii $r_i$ and
heights $h_i$ on thin wetting layers of thickness $w_i$ (see
Fig.~\ref{fig:scheme_molecule}). The QDs are centered at $\rho = 0$ of a
cylindrical coordinate system with the coordinates $(\rho,\phi,z)$. The
confinement potential for electrons and holes is given by the band edge
discontinuity of conduction and valence band, respectively, between the QD
and the host material. The confined electron (hole) states are calculated
within a variational multicomponent envelope function scheme as has been
presented in detail in Ref.~\onlinecite{Gawarecki2010}. In this scheme, an
``adiabatic'' approximation of the single-particle wave functions
$\psi^{l,M}(\rho,z,\phi)$ is used (cf. Ref.~\onlinecite{Wojs1996}). We use
the trial functions
\begin{equation}
\psi^{l,M}(\rho,z,\phi) = \frac{1}{\sqrt{2\pi}} \sum_{\alpha}
\chi_\alpha (z;\rho) \varphi^{l,M}_\alpha (\rho)\text{e}^{iM\phi},
\label{EQN:ansatz}
\end{equation}
where $M$ is the quantum number describing the projection of the angular
momentum on the $z$-direction, and $l$ is another quantum number that
uniquely labels the states with a given angular momentum. The functions
$\chi_\alpha (z;\rho)$ are the solutions of the one-dimensional Schr\"odinger
equation of the double well potential along the $z$-direction for a fixed
$\rho$,
\begin{equation}
\left\{ - \frac{\hbar^2}{2m_{\text{c/v}}} \frac{\partial^2}{\partial z^2} +
E_{\text{c/v}}(z;\rho) \right\} \chi_{\alpha}(z;\rho) = E_{\alpha}(\rho)
\chi_{\alpha}(z;\rho),
\end{equation}
representing the subbands of the confined states in the double well. Here,
$E_{\text{c/v}}(z;\rho) = V_{\text{c/v}}(z;\rho) + V(z)$  is the effective
confinement potential consisting of the confinement potential
$V_{\text{c/v}}(z;\rho)$ resulting from the band edge discontinuities of the
conduction/valence band and the potential $V(z)$ of the external electric
field.\cite{note2013} From the potential along the $z$-direction, we take the
lowest three confined states into account. Because the confinement potential
of the dots varies slowly with $\rho$, in Eq.~\fref{EQN:ansatz},
$\varphi^{l,M}_\alpha (\rho)$ is a slowly varying envelope function for this
coordinate. The lowest $\psi^{l,M}$-states can be obtained by applying Ritz'
variational method and minimizing the energy functional $F[\psi^{l,M}]$
within the class of trial functions defined in Eq.~\eqref{EQN:ansatz},
\begin{eqnarray}
F[\psi^{l,M}] &=& \langle \psi^{l,M} \left|H \right| \psi^{l,M} \rangle = \sum_{\alpha,
\beta} \int^{\infty}_0 \int^{\infty}_{-\infty} \rho \frac{\hbar^2}{2m} \nonumber\\ &&
\hspace{-1.2cm} \times \frac{\text{d}}{\text{d}\rho} \left( \chi_\alpha
(z;\rho)\varphi^{l,M}_\alpha (\rho) \right)^* \frac{\text{d}}{\text{d}\rho}\left(
\chi_\beta (z;\rho)\varphi^{l,M}_\beta (\rho) \right) \text{d}\rho \text{d}z \nonumber\\
&& \hspace{-1.2cm} +\sum_\alpha \int^{\infty}_0  \varphi^{l,M^*}_\alpha(\rho) \left(
E_{\alpha}(\rho) + \frac{\hbar^2 M^2}{2m\rho^2} \right)
\varphi^{l,M}_\alpha(\rho)\rho \text{d}\rho \nonumber \\
&& \hspace{-1.2cm} - \lambda \left( \sum_\alpha \int_0^\infty
\varphi_\alpha^{l,M^*}(\rho)\varphi_\alpha^{l,M}(\rho) \rho
\text{d}\rho - 1 \right),
\end{eqnarray}
where $\lambda$ is a Lagrange multiplier introduced to enforce normalization.
This minimization can be converted into an eigenvalue problem with $\lambda$
becoming the energy eigenvalues, i.e., $\epsilon^\text{e}_i$
($\epsilon^\text{h}_j$) for the electron (hole) energies, where $i$ ($j$) is
a multi-index including both quantum numbers $l$ and $M$.

The single-particle functions of electron and hole found in this way are used
to calculate the exciton states within a configuration interaction scheme.
For this purpose, we take the space spanned by the products of the
single-particle states, i.e. $\tilde{\Psi}_{ij} = \psi^{\text{e}}_i
\psi^{\text{h}}_j$, as a basis. Including the Coulomb interaction between
electron and hole, the Hamiltonian in this basis reads
\begin{eqnarray}
H_{\text{exc}}&=& \sum_{i} \epsilon^\text{e}_i c\dag_i c\ndag_i + \sum_{j}
\epsilon^\text{h}_j d\dag_j d\ndag_j + \sum_{ijkl} v_{ijkl} c\dag_i d\dag_j d\ndag_k
c\ndag_l,
\label{EQN:Coulomb}
\end{eqnarray}
where $c\ndag_i, c\dag_i$ denote the annihilation and creation operators for
the electron and $d\ndag_j,d\dag_j$ likewise for the hole, while
$\epsilon^\text{e}_i$ and $\epsilon^\text{h}_j$ are the corresponding
single-particle energies. The interaction between both particles is described
by the last term of the Hamiltonian, with
\begin{eqnarray}
v_{ijkl} &=& -\frac{e^2}{4 \pi \varepsilon_0 \varepsilon_r}\\ &&
\times \int \int {\psi_i^\text{e}}^*(\vec{r}) {\psi_j^\text{h}}^*(\vec{r}')
\frac{1}{|\textbf{r}-\textbf{r}^{'}|}{\psi_k^\text{h}}(\vec{r}') {\psi_l^\text{e}}
(\vec{r})\text{d}^3{r}\text{d}^3{r}^{'}, \nonumber
\end{eqnarray}
where $e$ is the elementary charge, $\varepsilon_0$ is the vacuum
permittivity, and $\varepsilon_r$ is the dielectric constant. This
Hamiltonian is diagonalized numerically within a truncated basis leading to
\begin{equation}
\label{ham-exciton}
H_{\text{exc}} = \sum_{n} \epsilon_n \ket{X_n}\bra{X_n} ,
\end{equation}
where the exciton energy $\epsilon_n$ corresponds to the state
\begin{equation}
\label{exciton-state}
\ket{X_n} = \sum_{ij} a^n_{ij} c\dag_i d\dag_j\ket{0},
\end{equation}
expressed as a superposition of the uncoupled electron-hole pair states with
expansion coefficients $a^n_{ij}$.

\subsection{Exciton-phonon coupling} \label{Theo:Phonon}

The excitons in the QDM are coupled to the phononic environment of the
structure. Because we are mainly interested in exciton states in an energy
range of less than 10~meV and we will consider only low temperatures, we can
restrict ourselves to the interaction with acoustic phonons which, due to the
very similar elastic properties of the QDs and the surrounding material, can
be treated as bulk-like. In QDs, due to the rather large splitting between
different confined energy levels, the coupling to acoustic phonons typically
gives rise to pure dephasing resulting in a strongly non-Lorentzian line
shape of absorption or emission lines \cite{Besombes2001,Krummheuer2002}
consisting of a narrow zero-phonon line (ZPL) superimposed on a broad phonon
background. Here, depending on the geometry of the QDM and the external
field, some exciton states may come energetically close such that transitions
between different excitonic states associated with the emission or absorption
of an acoustic phonon become possible, which gives rise to an additional
Lorentzian broadening of the ZPLs.

The coupling of the single carriers (electron and hole) to the phonon
environment can be described by the generic Hamiltonian
\begin{eqnarray}
H_{\text{c-ph}} &=& \sum_{s,\qu} \left( \sum_{ij} c\dag_i c\ndag_j
F^\text{e}_{s,ij}(\qu) + \sum_{kl} d\dag_k d\ndag_l F^\text{h}_{s,kl}(\qu)
\right)\nonumber \\&& \times \left( b\ndag_{s,\qu} + b\dag_{s,-\qu} \right).
\end{eqnarray}
Here, $F^{\text{e/h}}_{s,ij}(\qu)$ are the carrier-phonon coupling constants
between the single-particle states ($i,j$) and a phonon of branch $s$ and
momentum $\qu$ which is described by the boson operators $b\ndag_{s,\qu}$ and
$b\dag_{s,\qu}$. The carriers are coupled to transverse acoustic (TA) phonons
via the piezoelectric (PE) coupling and to longitudinal-acoustic (LA) phonons
via the PE and the deformation-potential (DP) coupling. The coupling
constants
$F^{\text{e/h}}_{s,ij}(\qu)=g^{\text{e/h}}_{s}(\qu)\mathcal{F}^{\text{e/h}}_{ij}(\qu)$
are composed of two parts. While the second term
\begin{eqnarray}
\mathcal{F}^{\text{e/h}}_{ij}(\qu) &=& \int {\psi^{\text{e/h}}_i}^*(\textbf{r})
\text{e}^{i\qu\cdot \textbf{r}} \psi^{\text{e/h}}_{j}(\textbf{r}) \text{d}^3r,
\label{coefficients}
\end{eqnarray}
is determined by the envelope functions of the involved states, the first
part $g^{\text{e/h}}_{s}(\qu)$ is the bulk coupling matrix element depending
on the specific mechanism. The bulk DP coupling constants are given by
\begin{eqnarray}
g^{\text{e/h}}_{\text{LA}}(\qu) &=& \sqrt{\frac{\hbar q}{2\rho V
v_\text{LA}}}a_{\text{c/v}},
\end{eqnarray}
where $\rho$ is the crystal density, $V$ is the system volume, $v_\text{LA}$
is the speed of sound of the LA phonons, and $a_{\text{c/v}}$ is a
deformation potential for the conduction or valence band. The piezoelectric
interactions can be described by the constants
\begin{eqnarray}
g^{\text{e/h}}_{s}(\qu) &=& -i
\sqrt{\frac{\hbar}{2\rho V v_s q}}\frac{d_p e}{\epsilon_0 \epsilon_r}
M_s(\hat{\qu}),
\end{eqnarray}
with $v_s$ being the speed of sound of the LA
and the TA phonons, respectively, while $d_p$ is the piezoelectric constant.
$M_s(\hat{\qu})$ describes the dependence of the PE coupling on the direction
$\hat{\qu} = \frac{\qu}{q}$ of the phonon wave-vector and is given by
\begin{eqnarray}
M_s(\hat{\qu}) &=& 2\left[\hat{q}_x(\hat{e}_{s,\qu})_y \hat{q}_z
+ \hat{q}_y(\hat{e}_{s,\qu})_z \hat{q}_x + \hat{q}_z(\hat{e}_{s,\qu})_x
\hat{q}_y\right],\qquad
\end{eqnarray}
where $\hat{e}_{s,\qu}$ is the unit polarization vector for the phonon branch
$s$ and the wave vector $\qu$.

The interaction of an exciton with the phonon environment is given by the
projection of $H_{\text{c-ph}}$ on the exciton states
\begin{eqnarray}
H_{\text{x-ph}} &=& \sum_{nm} \sum_{s,\qu} F_{s,nm}^\text{x}(\qu) \left(
b\ndag_{s,\qu} + b\dag_{s,-\qu} \right) \ket{X_n}\bra{X_m},\quad
\end{eqnarray}
with the exciton-phonon coupling constant
\begin{eqnarray}
F_{s,nm}^\text{x}(\qu) &=& \sum_{ij}
\bra{X_n} c\dag_i c\ndag_{j} \ket{X_m} F^\text{e}_{s,ij}(\qu) \nonumber\\
&&+ \sum_{kl} \bra{X_n} d\dag_k d\ndag_{l} \ket{X_m} F^\text{h}_{s,kl}(\qu).
\end{eqnarray}
Since the exciton states are superpositions of electron-hole configurations,
this may give rise to transitions between exciton states. The transition rate
$\gamma_{nm}$ for a transition between an initial exciton-state $n$ and a
final state $m$ with energies $E_n$ and $E_m$, respectively, is given by
Fermi's golden rule
\begin{eqnarray}
\gamma_{nm} =
\frac{2 \pi}{\hbar^2} \left[ n_\text{B}(T, \omega_{nm}) + \delta \right]
J_{nm}(\omega_{nm}),
\label{gamma_if}
\end{eqnarray}
where $\Delta E_{nm} = \hbar \omega_{nm} = |E_n - E_m|$, $n_\text{B}(T,
\omega_{nm})$ is the Bose distribution at the lattice temperature $T$, and
$\delta = 0$ in case of a phonon absorption (i.e., $E_m > E_n$) and $\delta =
1$ in case of a phonon emission (i.e., $E_m < E_n$). The phonon spectral
density $J_{nm}(\omega)$ for the transition from state $n$ to state $m$ is
defined as
\begin{eqnarray}
J_{nm}(\omega) &=& \sum_{s,\qu}\delta(\omega -
\omega_{s,\qu})\left| F_{s,nm}^\text{x}(\qu) \right|^2.
\end{eqnarray}
The transition rate between two states depends on the applied field, as the
energy difference and the wave functions of the involved states are affected
by the field. The total decay rate $\gamma_n$ of an exciton state $n$ is
given by the sum over all possible transitions, $\gamma_n=\sum_m
\gamma_{nm}$. The lifetime $\tau_n$ of the state is then defined by the
inverse of the transition rate, $\tau_n = \gamma_n^{-1}$.

\subsection{Absorption Spectrum} \label{Theo:Absorp}
To investigate the absorption spectrum of the double quantum dot system we
consider the dipole coupling of the carriers to a classical light field
within the rotating wave approximation\cite{Rossi2002,Binder1994}
\begin{equation}
H_{\text{c-l}} = - \sum_{ij} \vec{M}_{i j} \cdot
\vec{E}^{(-)} c\ndag_i d\ndag_j + \text{h.c.},
\end{equation}
where $\vec{E}^{(-)}$ is the negative frequency part of the field and
$\vec{M}_{i j}$ is the dipole matrix element of the interband transition,
which is given by the product of the bulk dipole matrix element $\vec{M}_0$
and the overlap integral between electron and hole envelope wave functions.
The linear absorption spectrum is proportional to the imaginary part of the
linear susceptibility $\chi(\omega)$,\cite{Fox} it can therefore be obtained
from the projection of the dipole operator on the direction of the light
field ${P}=\sum_{i j} {M}_{i j} c\ndag_i d\ndag_j = \sum_{n} {M}_{n}
\ket{0}\bra{X_n} = \sum_{n} {M}_{n} a_n $, where $a_n$ are the exciton
annihilation operators and ${M}_{n}$ the corresponding dipole matrix
elements. For a weak, ultra-fast pulse (linear response) at $t =0$, i.e.,
$E(t)=E_0 \delta(t)$ one gets \cite{Suna1964,Jacak2005,Jacak2005_2}
$\bks{\bks{P(t)}} \sim \chi(t) E_0$ with
\begin{eqnarray}
\chi(t) &\sim& \Theta(t) \sum_{n m} {M}^{ }_{n}
{M}^*_{m} \bks{\bks{ a\ndag_n(t) a\dag_m }} \nonumber \\ &=& \sum_{n m}
{M}^{ }_{n} {M}^*_{m} G_{n,m}(t).
\end{eqnarray}
Here, $\bks{\bks{\dots}}$ is the grand canonical average, which is
temperature dependent because of the phonon degrees of freedom. We calculate
the polarization by utilizing the single-particle Green's function
$G_{n,m}(\omega) = \int^{\infty}_{-\infty} G_{n,m}(t)e^{i \omega t}\dt$. Up
to the second order in the phonon coupling and neglecting phonon-induced
energy shifts $G_{n,m}(\omega)$ can be approximated by\cite{Jacak2005_2}
\begin{equation}
G_{n,m}(\omega) = \frac{ \delta_{n,m} }{ \omega -
\frac{\epsilon_n}{\hbar} + i \tilde{\gamma}_n(\omega)}
\end{equation}
with the imaginary part of the self-energy
\begin{equation}
\tilde{\gamma}_n(\omega) = \sum_{l}\tilde{\gamma}_{nl}(\omega)
\end{equation}
and
\begin{eqnarray} \tilde{\gamma}_{nl}(\omega) &=&\frac{\pi}{\hbar^2} \sum_{s,\qu}
\left| F_{s,nl}^\text{x}(\qu) \right|^2 \nonumber\\&& \times \left\{
n_\text{B}(T, \omega_{s,\qu}) \delta\left( \omega - \frac{\epsilon_l}{\hbar}
+ \omega_{s,\qu}\right) \right. \label{gamma}
\\  &&\left.+ \left[n_\text{B}(T, \omega_{s,\qu}) + 1\right]
\delta\left(\omega - \frac{\epsilon_l}{\hbar} - \omega_{s,\qu}\right)
\right\},\nonumber
\end{eqnarray}
being the contribution to the self-energy resulting from transitions between
the states $n$ and $l$. The off-diagonal  Green's function $G_{n,m}(\omega)$
with $n \neq m$ is of higher order in the phonon coupling and has therefore
been neglected. Thus, for the absorption spectrum $I(\omega)$ we get
\begin{eqnarray}
I(\omega) &\sim& \textrm{Im} \chi(\omega) \sim \sum_n |{M}^{ }_{n}|^2 \frac{
\tilde{\gamma}_n(\omega) }{ \left(\omega - \frac{\epsilon_n}{\hbar}\right)^2
+ \tilde{\gamma}_n(\omega)^2}.
\end{eqnarray}
This expression accounts for spectral broadening due to phonon-mediated
transitions as well as pure dephasing.

The off-diagonal contributions $\tilde{\gamma}_{nl}(\omega)$ ($l \neq n$)
result from real phonon-mediated transitions from the exciton states $n$ to
the state $l$. Here the frequency can be replaced by the value of the initial
state, i.e., $\omega=\epsilon_n/\hbar$. Formally this corresponds to a Markov
approximation equivalent to the derivation of Fermi's golden rule. Indeed,
when comparing with Eq.~\eqref{gamma_if} we have $\tilde{\gamma}_{nl} =
\frac{1}{2} \gamma_{nl}$. Including only off-diagonal contributions therefore
results in an absorption line with a Lorentzian broadening caused by the
finite lifetime of the exciton state due to phonon-mediated transitions to
other exciton states.

The diagonal part $\tilde{\gamma}_{nn}(\omega)$, on the other hand, is not
related to phonon-mediated transitions between different exciton states,
instead it describes the pure dephasing contribution of the state $n$ to the
spectrum. The pure dephasing contribution vanishes in the Markov limit
because the diagonal form factor $F_{s,nn}^\text{x}(\qu)$ vanishes for $q \to
0$. Therefore, pure dephasing is a genuine non-Markovian process where the
full frequency dependence has to be kept. In the spectrum, it causes the
presence of phonon sidebands next to the resonant excitation of the state $n$
(zero phonon line). Indeed, when reducing the system to a single exciton
state Eq.~\fref{gamma} agrees with the perturbative result for pure dephasing
as derived in Ref.~\onlinecite{Krummheuer2002}.

Including both diagonal and off-diagonal contributions to
$\tilde{\gamma}_{n}$ the spectrum therefore consists of a zero phonon lined
broadened by real transitions to other exciton states and phonon sidebands
describing optical transitions assisted by the emission or absorption of a
phonon. For small QDs the pure dephasing contribution is particularly
important due to the large separation in energy of excitonic states and the
resulting suppression of real phonon-mediated transitions. The off-diagonal
part $\tilde{\gamma}_{nl}(\omega)$, on the other hand, becomes important for
a system with exciton states which are energetically close, such that they
can be reached by the emission or absorption of an acoustic phonon which is
the case, e.g., in QDMs or larger QDs.

Besides phonon-mediated transitions to other exciton states also other
processes which limit the lifetime of an exciton state contribute to the
broadening of the zero phonon line. These are in particular the radiative
decay of the exciton or the tunneling of electron and hole out of the QD, if
the QD is embedded in a photodiode structure. To model such processes, in our
calculations we add a finite background lifetime, for which we take a value
of 250~ps. In higher orders of the carrier-phonon coupling also additional
phonon-related processes might contribute, such as two-phonon scattering via
a virtual excited state. \cite{Muljarov2005} These effects are neglected
here.

If we consider photoluminescence instead of absorption, the linewidths caused
by real phonon-mediated transitions are essentially the same. However, the
sidebands on the high and low energy side of the zero phonon lines (ZPLs)
resulting from phonon-assisted optical transitions are mirrored. This is
because phonon emission gives rise to the low energy sideband in
photoluminescence, but to the high energy sideband in absorption, and vice
versa for phonon absorption. Here we will concentrate on absorption spectra
as can be measured, e.g., by using photocurrent spectroscopy because this
technique gives access not only to the lowest exciton state but also to
higher states. In contrast, photoluminescence signals, in particular at low
excitation intensities, are often restricted to the lowest exciton states.

\subsection{Spin-Orbit Coupling} \label{Theo:SOC}
The considered semiconductors, InAs and GaAs, have a zincblende structure,
characterized by a bulk inversion asymmetry (BIA).\cite{Dresselhaus55}
Therefore, even for a cylindrically symmetric nanostructure, SOC couples the
spins of electron ($\uparrow$, $\downarrow$) and hole ($\Uparrow$,
$\Downarrow$) to the orbital angular momenta.\cite{Winkler03} For electrons,
we consider the Dresselhaus term which is of third order in the wave vector
$\ka$, \cite{Winkler03}
\begin{eqnarray}
H^{\text{e}}_{\text{SOC}} &=&
b^{\text{6c6c}}_{41} \left\{  \Big[\frac{k_{-}}{2}\big( k_{+}^2 - k_{-}^2 \big)
-2k_{+}k_{z}^2 \Big]\sigma_{+} +\text{h.c.} \right\}\nonumber \\
&& + b^{\text{6c6c}}_{41} k_z \big( k_{+}^2 + k_{-}^2 \big)\sigma_z,
\end{eqnarray}
with $k_{\pm} = k_{x}\pm i k_{y}$ and $\sigma_{\pm} = \frac{1}{2}(\sigma_{x}
\pm i \sigma_{y})$, where $\sigma_i (i=x,y,z)$ are the Pauli matrices. With a
restriction to heavy holes, the BIA spin-orbit coupling for holes
reads,\cite{Winkler03}
\begin{equation}
H^{\text{h}}_{\text{SOC}} = \begin{pmatrix} H_{\Uparrow } & H_{\Uparrow \Downarrow} \\
H_{\Downarrow \Uparrow} & H_{\Downarrow } \end{pmatrix}
\end{equation}
with
\begin{eqnarray}
H_{\Uparrow} &=& \frac{27}{8} b^{\text{8v8v}}_{42}\big( k_{+}^2 + k_{-}^2 \big) k_{z}
= H_{\Downarrow}
\end{eqnarray}
and
\begin{eqnarray}
H_{\Uparrow \Downarrow} &=& -\frac{1}{8}\Big[ \big(24 b^{\text{8v8v}}_{51} + 12
b^{\text{8v8v}}_{42} \big) k_{-}k_{z}^2 + 8\sqrt{3}C_{k}k_{-} \qquad \nonumber \\
&&  +\big(9 b^{\text{8v8v}}_{52} + 6 b^{\text{8v8v}}_{51} - 3 b^{\text{8v8v}}_{42} \big)
k_{-}^2k_{+} \nonumber \\
&&  +\big(3 b^{\text{8v8v}}_{52} - 6 b^{\text{8v8v}}_{51} + 3
b^{\text{8v8v}}_{42} \big) k_{+}^3 \Big] = H\dag_{\Downarrow \Uparrow}.
\end{eqnarray}
Although this interaction is typically weak, it provides a coupling between
states with different angular momenta which is absent otherwise. We will
therefore take spin-orbit coupling into account only for specific situations,
where a qualitative difference is expected due to this coupling (in
particular, near crossings of two otherwise uncoupled levels), and neglect it
otherwise. In these situations, we need to consider the electron-hole
exchange interaction in the same dot as well, as the exchange interaction is
of comparable strength,
\begin{equation}
H_{\text{Ex}} = j_{\text{eh}}\vec{S}^{\text{e}}\cdot\vec{S}^{\text{h}},
\end{equation}
where $\vec{S}^{\text{e}}$ and $\vec{S}^{\text{h}}$ are the spin operators of
electron and hole. For InGaAs QDs this leads to a splitting between bright
and dark exciton spin-configurations of about
100~$\mu$eV.\cite{Bayer2002,Seguin2005}

In principle, we have to consider the $\ka$-linear Rashba effect as well,
because the QD structure itself as well as the applied electric field
$\vec{F}= -F \vec{e}_z$ also introduce a structure inversion asymmetry (SIA).
For the heavy holes, this leads to a coupling\cite{Winkler03}
$H^{\text{r}}_{\Uparrow \Downarrow} = \frac{3 i}{4} r^{\text{8v8v}} F k_{+} =
H^{\text{r}\dagger}_{\Downarrow \Uparrow}$ and $H^{\text{r}}_{\Uparrow} =
H^{\text{r}}_{\Downarrow} = 0$, while for the electrons this
gives\cite{Winkler03} $H^{\text{r}}=i r^{\text{6c6c}}_{41} F \left( k_{-}
\sigma_{+} - k_{+} \sigma_{-}  \right)$. However, for the electric fields $F$
considered here, these couplings can be estimated to be three and two
orders of magnitude smaller, respectively, compared to the similar
Dresselhaus terms. We will therefore neglect the Rashba effect.

\section{Results} \label{Res}
\begin{table}
    \centering
    \caption{Geometry of the quantum dot molecule.}
        \begin{tabular} { l c  c  c  } \hline \hline
                                &       &   lower dot   & upper dot \\ \hline
  base radius (nm)      & $r_i$ &  12.0         &  16.0     \\
  center height (nm)    & $h_i$ &   4.4         &   3.8     \\
  wetting layer (nm)    & $w_i$ &   0.5         &   0.5     \\
  In content            & $x$   &   0.3         &   0.3   \\
  spacer layer (nm)     & $D$   &    \multicolumn{2}{c}{9.9}  \\ \hline \hline
        \end{tabular}
    \label{table1}
\end{table}
In this section, we present exemplary calculations for a semiconductor QDM.
The geometry parameters are summarized in Table \ref{table1}, they are in the
parameter range of typical strain-induced InGaAs
QDMs.\cite{Krenner2005,Mueller2011,Szafran2008} The material parameters that
have been used in the calculations are given in Table \ref{table2}. In
Secs.~\ref{Res:ExSt} - \ref{Res:Phonon} we neglect SOC and assume a structure
with a perfect cylindrical symmetry. The role of SOC and the breaking of the
cylindrical symmetry will then be discussed in Secs.~\ref{Res:SOC} and
\ref{Res:Broken}.
\begin{table}[t]
    \centering
    \caption{Material parameters for carrier wave-functions, phonons and SOC used in the calculations. }
        \begin{tabular} { l c  r  } \hline \hline
  \multicolumn{3}{r}{In$_{0.3}$Ga$_{0.7}$As}                        \\\hline
  \multicolumn{3}{l}{electronic parameters}                         \\
  electron mass ($m_0$)                 & $m_\text{e}$  &  0.09     \\
  hole mass ($m_0$)                     & $m_\text{h}$  &  0.37     \\
  relative dielectric constant          & $\epsilon_r$  &  12.9     \\
  band edge discontinuities (\text{eV}) & $E_\text{c}$  & -0.255    \\
                                        & $E_\text{v}$  &  0.075    \\
  exchange interaction (\text{meV})      & $j_{eh}$     & -0.133    \\
  \multicolumn{3}{l}{phonon parameters}  \\
  speed of sound (\text{m/s})           & $v_\text{LA}$ &  5110     \\
                                        & $v_\text{TA}$ &  3340     \\
  crystal density (\text{kg/m}$^3$)     & $\rho$        &  5300     \\
  deformation potentials (\text{eV})    & $a_\text{c}$  & -6.90     \\
                                        & $a_\text{v}$  &  2.65     \\
  piezoelectric constant (\text{C/m}$^2$)& $d_p$        & -0.16     \\
  \multicolumn{2}{l}{spin-orbit coefficients}           &           \\
  \multicolumn{2}{l}{ $b_{41}^{6c6c}$ (eV\r{A}$^3$)}    & 27.46     \\ 
  \multicolumn{2}{l}{  $b_{51}^{8v8v}$ (eV\r{A}$^3$)}   &  0.469    \\
  \multicolumn{2}{l}{  $b_{42}^{8v8v}$ (eV\r{A}$^3$)}   &  1.407    \\
  \multicolumn{2}{l}{  $b_{52}^{8v8v}$ (eV\r{A}$^3$)}   & -0.938    \\
  \multicolumn{2}{l}{  $C_k$ (eV\r{A}) }                & -0.00574\\ \hline \hline
        \end{tabular}    \label{table2}
\end{table}

\subsection{Exciton States} \label{Res:ExSt}

\begin{figure}[tb]
    \includegraphics[width=1.0\columnwidth]{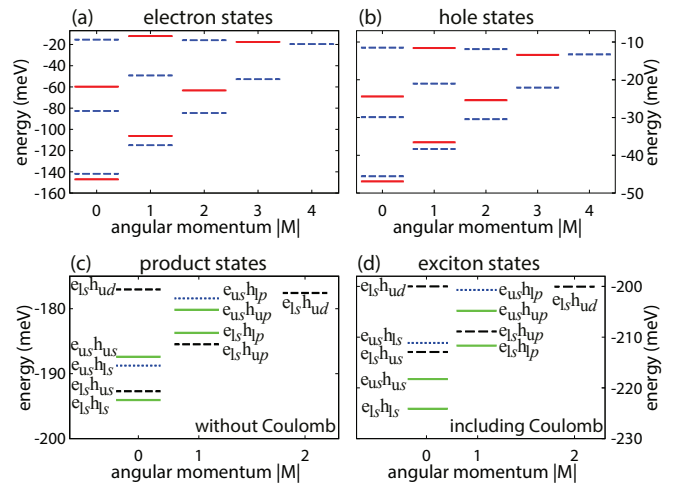}
    \caption{(Color online) Energy-level structure in dependence of angular momentum $M$ in
    the absence of an external electric field. (a)
    Electron states with main localization in lower (red, solid) or upper
    (blue, dashed) dot. (b) Hole states in the same way. (c)
    Product states of electron and hole (without Coulomb interaction),
    classified according to a spatially direct configuration (green, solid) and
    an indirect configuration with a positive (black, dashed) or negative
    (blue, dotted) sign of the electric dipole moment. (d) Exciton states
    (including Coulomb interaction) with the same coding as in (c). The labels
    in (c) and (d) refer to the main localization of electron (e) and hole (h) in
    the upper (u) or lower (l) dot as well as the shell ($s,p,d$) of the
    dominant contribution.
    } \label{states}
\end{figure}

Figure~\ref{states} shows the energy levels of the confined electron (a),
hole (b), product (c) and exciton (d) states. Electrons and holes exhibit
essentially the same structure, but the absolute values and separations of
the hole energies are much smaller because of the weaker confinement and the
larger mass. For the chosen geometry of the QDM, the lowest state of both
carriers is mainly localized in the lower dot (red, solid lines), which is,
in terms of the standard terminology, of an $s$-shell like character. This
state is closely followed by another $s$-shell like state localized in the
upper dot (blue, dashed lines). This sequence is caused by the larger height
of the lower dot. The energetically next higher state of each dot has an
angular momentum of one ($p$-shell). Due to the cylindrical symmetry of the
system, the $p$-shells of each dot consist of two degenerate states with
$M=\pm 1$. For the $p$-shells, as for all subsequent shells, the lowest state
is localized in the upper dot. The order in energy is inverted compared to
the $s$-shell because the base radius of the upper dot is larger than the
radius of the lower dot. Hence, the almost equidistant sequence of shells of
the upper dot is more compact than for the lower dot. For a $d$-shell state
we have to distinguish between an angular momentum of two and zero. These
states are not degenerate, as we do not have a harmonic confinement (for the
hole: $E_{M=0} - E_{M=2} \approx 1~$meV). In the following, we denote the
single-particle states by their main localization (u for upper and l for
lower dot) and their basic type ($s$-, $p$-, $d$-\dots shell like state),
e.g., e$_{\text{l}s}$ for an electron $s$-shell state in the lower dot.

From the electron and hole single-particle states pair states (excitons) can
be formed. For such pair states in a QDM, one has to distinguish between a
spatially direct and a spatially indirect configuration. The electron and the
hole of a direct exciton are mainly localized in the same dot. The behavior
of such an exciton is similar to an exciton in a single QD. In case of an
indirect configuration, the electron and the hole are mainly localized in
different dots. Therefore, even without an applied electric field the
indirect exciton has an electric dipole moment $\vec{p}$, roughly
proportional to $\pm e\,D$. Thus its energy shifts linearly with an external
electric field $\vec{F}$. Because the overlap of the electron and hole wave
functions is negligible, indirect excitons are not optically active. However,
they can become visible in an absorption spectrum due to state mixing with
bright, direct excitons, if the coupling between these two configuration is
allowed.

In Fig.~\ref{states}(c) we have plotted a section of the energy structure of
the product states of electron and hole ($\tilde{\Psi}_{ij}$, neglecting
Coulomb interaction). The states are distinguished according to a direct
configuration (green, solid lines), an indirect configuration with a positive
dipole moment (black, dashed lines), and an indirect configuration with a
negative dipole moment (blue, dotted lines).

\begin{figure*}[ht]
    \includegraphics[width=1.95\columnwidth]{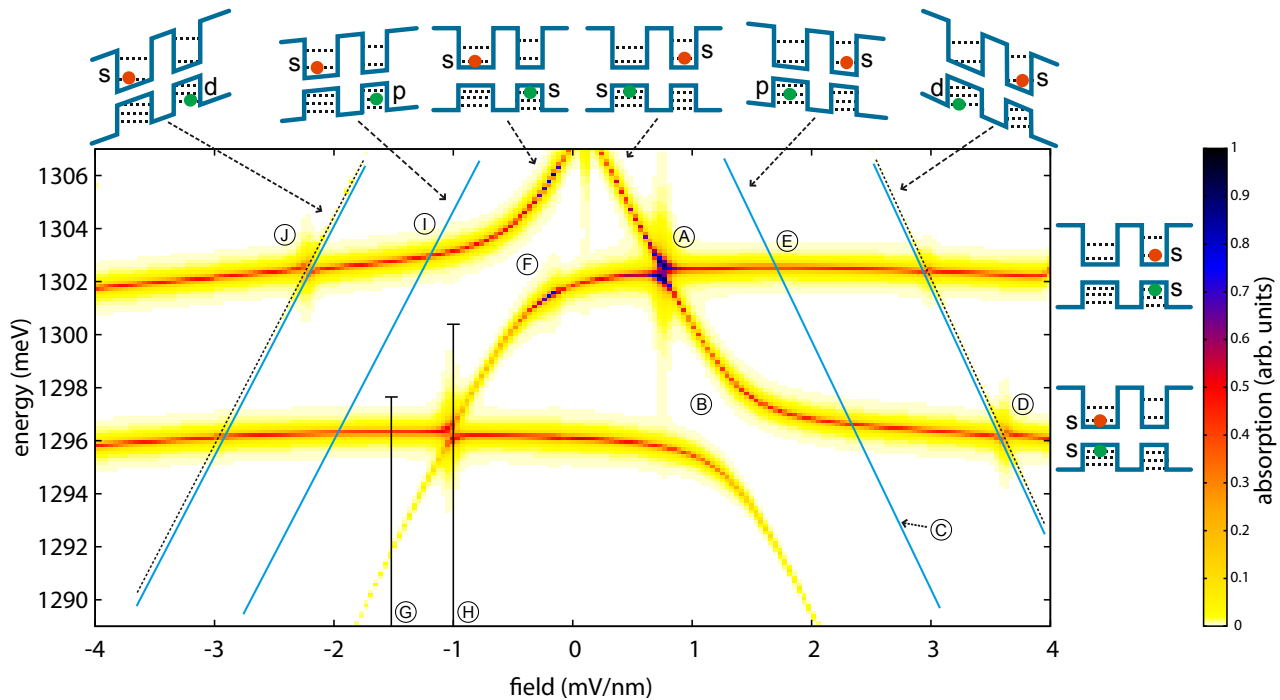}
    \caption{(Color online) Absorption spectrum as a function of the applied electric field at
    a temperature of $T=10$~K with schemes of the involved exciton states. In
    the schemes the right potential well refers to the upper dot and the left
    well to the lower dot. The labels refer to the spectral features and
    positions discussed in the text. Vertical bars at \Tag{G} and \Tag{H} indicate the
    spectral range shown in Fig.~\ref{puredeph}.}
    \label{Absorp_LS}
\end{figure*}

Figure~\ref{states}(d) shows the energies of the exciton states including
Coulomb interaction, as obtained from the diagonalization of the Hamiltonian
in Eq.~\fref{ham-exciton}. The coding of the lines is the same as in part
(c). Although an exciton state is a superposition of electron-hole product
states according to Eq.~\fref{exciton-state}, there is in many cases a
dominant electron- and hole-contribution, which we use to label the exciton
state, e.g., as \X{\ls}{\up}. In the absence of Coulomb coupling, the lowest
indirect states, \X{\us}{\ls} and \X{\ls}{\us}, are lower in energy than the
second direct state \X{\us}{\us}. This is reversed by taking Coulomb
interaction into account, because of the much stronger attraction of electron
and hole for a direct exciton. Besides these localization-dependent energy
shifts, Coulomb interaction couples different product states, which is
important to understand some phenomena of the absorption spectra in the next
section.

\subsection{Absorption spectra} \label{Res:Absorbtion}

Figure~\ref{Absorp_LS} shows a contour plot of the absorption spectrum of the
QDM as a function of the photon energy $E$ and the electric field $F$, which
is applied opposite to the growth direction, i.e., a positive value of $F$
refers to a positively charged top contact. The spectral widths of the lines
caused by the interaction with phonons can be extracted from the color
coding.

The horizontal line at 1.296~eV corresponds to the lowest direct exciton
state localized in the lower dot, \X{\ls}{\ls}, while the line at 1.302~eV
corresponds to the direct exciton in the upper dot, \X{\us}{\us}. For the
given values of QD heights and applied fields, the quantum confined Stark
effect (QCSE) is only slightly visible by the slight curvature of the
absorption lines of the direct excitons with a maximum at zero field. The
horizontal lines are intersected by lines corresponding to indirect exciton
states, the energy of which shifts linearly with the applied field due to the
linear Stark effect $E_{\text{Stark}} \propto - \vec{p}\cdot\vec{F} \approx
\pm eDF$. Hence, for increasing electric fields $F$, indirect-exciton states
with the hole mainly localized in the lower dot and the electron mainly
localized in the upper dot [blue, dotted states in Fig.~\ref{states}(d)]
shift downward in energy, while those with the opposite localization of
electron and hole [black, dashed states in Fig.~\ref{states}(d)] shift upward
in energy. The former states have the following sequence in energy:
\X{\us}{\ls}, \X{\us}{\lp}, \X{\us}{\ld}, \X{\us}{\lf}, \X{\up}{\ls}, \dots
while the latter have the same sequence with lower and upper dot exchanged.
This order holds for a wide range of parameters, while the distance in energy
depends on the specific choice of parameters, like the ratio of carrier
masses or strengths of confinement potentials.

As is evident from Fig.~\ref{Absorp_LS}, at some of the intersections between
direct and indirect excitons the energy levels just cross without being
visible in the absorption spectrum while at other points a more or less
pronounced avoided crossing appears in the spectrum. At \Tag{A} we observe
such an avoided crossing of the direct state \X{\us}{\us} and the indirect
state \X{\us}{\ls}. Here the hole states in the two dots mix due to tunnel
coupling which makes both excitons optically bright in the mixing region. An
even more pronounced avoided crossing of \X{\ls}{\ls} and \X{\us}{\ls}
appears at \Tag{B}, which is induced by the tunnel coupling of the electron
states in the two dots. The difference in strength of these avoided crossings
stems mainly from the difference of electron and hole masses, which favors
electron tunneling.

The indirect exciton with the hole having the character of a $p$-shell state
(\Tag{C}; blue, solid line) runs about 11~meV above the previous
indirect-exciton state. Due to conservation of angular momentum this state is
dark (dipole-forbidden) and decoupled from the visible direct-exciton states.
A straight crossing with both direct states occurs, which however can be
changed by SO coupling (cf.~Sec.~\ref{Res:SOC}) or a perturbation of the
cylindrical symmetry of the system (cf.~Sec.~\ref{Res:Broken}).

For an exciton with a hole in the $d$-shell (\X{\us}{\ld}), there are three
different configurations: the dark, decoupled ones with angular momenta
$M=\pm 2$ (solid, blue line), and the dipole-allowed configuration with $M=0$
(dotted, black line). The former states are not coupled to the direct state
\X{\ls}{\ls} due to the conservation of angular momentum. But also the latter
one is not coupled by tunnel coupling to this direct state because tunneling
is a single-particle process, while in these two pair-configurations the
states of both particles differ. Hence, in the absence of Coulomb
interaction, straight crossings would occur. However, a weak anticrossing at
a field of $3.6$~mV/nm, \Tag{D}, appears between the two states with $M=0$
because of Coulomb interaction. Being a a two-particle interaction, it leads
to a coupling between excitons with the same total angular momentum,
especially at an exciton resonance, i.e., E(\X{\us}{\ld}) $\approx$
E(\X{\ls}{\ls}), even if all four involved single-particle states are
different. The strict discrimination between the $d$-shell states with $M=0$
and $M=\pm 2$ is a consequence of the cylindrical symmetry, which is why in
the case of a broken symmetry the anticrossing structure becomes more
complicated (cf.~Sec.~\ref{Res:Broken}).

The next two indirect states (\X{\us}{\lf} and \X{\up}{\ls}) have a nonzero angular
momentum. They have no dipole-allowed transitions and are decoupled from the direct
states with angular momentum $M=0$, and are therefore not depicted.

For negative fields the indirect states with a negative dipole moment shift
up in energy, while the indirect states with the hole mainly localized in the
upper dot (black, dashed lines in Fig.~\ref{states}(d)) shift down. Hence,
the upper anticrossing at $-0.5$~mV/nm (\Tag{F}) is caused by the tunnel
coupling of the electron in \X{\us}{\us} and \X{\ls}{\us}, instead of the
hole-tunnel coupling for positive fields at \Tag{A} with \X{\us}{\ls}.
Basically the sequence of crossings and anticrossings of upper and lower dot
swaps with the sign of the applied field (\Tag{F} corresponds to \Tag{B},
\Tag{J} corresponds to \Tag{D} and so on). In this case however, the
distances in field strength between the crossings and anticrossings are
smaller due to the smaller energy distance of the upper dot's states. This
sequence of crossings and anticrossings is similar to the case of positive
field strengths but with upper and lower dots interchanged (although minor
differences occur, as the QDM is not symmetric with respect to the growth
axis).

\subsection{Phonon-mediated transitions} \label{Res:Phonon}

\begin{figure}
    \includegraphics[width=0.99\columnwidth]{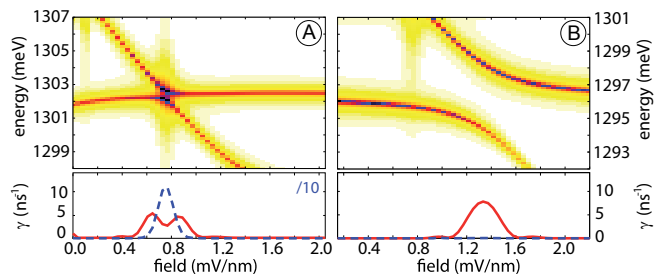}
    \caption{(Color online) Absorption spectrum as a function of the applied electric field
    (top) and phonon-mediated transition rates from the upper to the lower state
    (bottom) due to piezoelectric
    coupling (blue, dashed) and deformation potential (red, solid) at 10~K. In the left panel the rate for
    piezoelectric coupling has been scaled down by a factor of $1/10$. Left:
    Hole-anticrossing \Tag{A}. Right: Electron-anticrossing \Tag{B}.}
    \label{AC1}
\end{figure}
In Fig.~\ref{Absorp_LS} the width of the absorption lines caused by the
interaction with acoustic phonons has already been indicated by the color
coding. We will now discuss these line widths in some more detail. In the
upper panels of Fig.~\ref{AC1} the absorption spectra in the region of the
anticrossings caused by the tunnel-coupling of the holes (\Tag{A}, left
figure) and of the electrons (\Tag{B}, right figure) are plotted again as
functions of the electric field. The lower panels show the corresponding rate
$\gamma$ for a transition from the upper to the lower state by phonon
emission due to DP (red, solid) and PE coupling (blue, dashed). Note that the rate for PE
coupling has been scaled down by a factor of $1/10$. Only in the regions of
the anticrossings there is a considerable overlap of the exciton states on
the different branches, such that phonon-assisted transitions can take place.
In the case of the hole-anticrossing, which is characterized by a weak
splitting, we find a large transition rate of the order of 100~ns$^{-1}$
corresponding to a lifetime of the upper state of about 10~ps due to the
piezoelectric coupling (blue, dashed line). This short lifetime is in good agreement
with recent experimental observations.\cite{Mueller2012} The PE coupling is
strong only in a small energy window of about $0.3 - 0.9$~meV, where it
strongly dominates over the DP coupling. DP coupling is typically strongest
for energy differences in the range of $1 - 2$~meV. Therefore the maxima of
the transition rate due to DP interaction in the case of the hole
anticrossing appear on both sides of the anticrossing. For the electron
anticrossing, the PE coupling is unimportant because of the large energy
difference of the states. Here the rate due to DP coupling has its maximum at
the center of the anticrossing. In both cases the rate reaches at most a
value of about 10~ns$^{-1}$ corresponding to a lifetime of about 100~ps. In
a QDM there is an oscillatory contribution to the transition rate, which is a
typical behavior for quasi one-dimensional phonon
scattering;\cite{Bockelmann1990,Gawarecki2010,Wijesundara2011, Rolon2012} it
is a result of a resonance condition between the phonon wavelength and the
distance between the dots.  This is the origin of the weak revivals on both
sides of the maximum.

\begin{figure}
    \includegraphics[width=0.99\columnwidth]{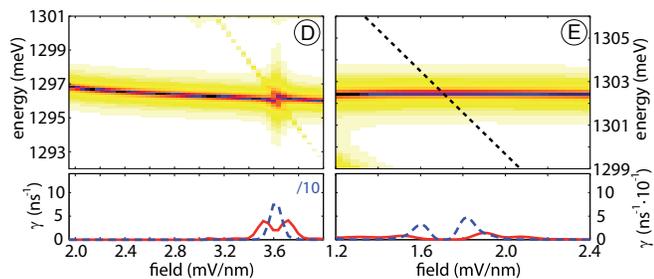}
    \caption{(Color online) Absorption spectrum as a function of the applied electric field (top) and
    phonon-mediated transition rates from the upper to the lower state
    (bottom) due to piezoelectric coupling (blue, dashed) and
    deformation potential (red, solid) at 10~K. In the left panel the rate for
    piezoelectric coupling has been scaled down by a factor of $1/10$.
    Left: Coulomb-anticrossing at \Tag{D}. Right:
    Crossing of exciton states with different angular momenta at \Tag{E}.}
    \label{AC2}
\end{figure}

Figure~\ref{AC2} shows in the same way the spectra in the region of the
Coulomb-induced exciton-anticrossing (\Tag{D}, left figure) and of a crossing
(\Tag{E}, right figure) of exciton states with different angular momenta of
$M=0$ (horizontal) and $|M|=1$ (dashed line). The strength of the Coulomb
anticrossing as well as the rates of the phonon-mediated transitions are
similar to the case of the hole-anticrossing in Fig.~\ref{AC1} (left). This
illustrates that the exciton transitions due to phonons do not depend on the
coupling mechanism between the involved states. They just depend on the
energy separation and the spatial overlap of the states. In the case of
crossing exciton states (Fig.~\ref{AC2}, right) phonon-assisted transitions
are almost absent. The involved states are completely uncoupled due to the
conservation of angular momentum and no mixing of the states occur. Since the
phonon modes do not have a well-defined angular momentum there is a slight
coupling resulting from the weak overlap of the wave functions, however the
rate remains very small.

\begin{figure}
    \includegraphics[width=0.99\columnwidth]{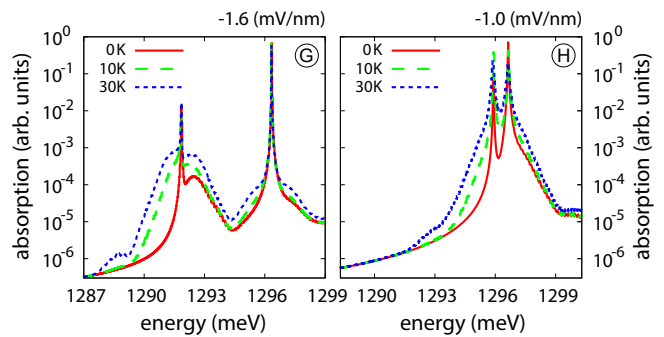}
    \caption{(Color online) Line shape of the absorption spectra for different temperatures at two
    values of the
    static electric field: At the hole tunnel-anticrossing \Tag{H} (-1~mV/nm)
    and at a lower field of -1.5~mV/nm \Tag{G}.}
    \label{puredeph}
\end{figure}

The shape of an absorption line depends crucially on the phonon environment
and therefore on temperature.\cite{Besombes2001,Borri2001,Borri2003} This is
illustrated in Fig.~\ref{puredeph} where absorption spectra for two fixed
electric fields are shown: At the hole tunnel-anticrossing (\Tag{H}, right
figure) and at a lower electric field on the same direct exciton line
(\Tag{G}, left figure). In each figure, two pronounced peaks (ZPLs) with a
temperature dependent background (sidebands) stemming from the states
\X{\ls}{\ls} and \X{\ls}{\us} are visible. This shape is characteristic for
pure dephasing in QDs. The central peaks are of Lorentzian shape with a
broadening that reflects the lifetime of the given state ($\tau_n$) with
respect to real transitions (relaxation processes). At low temperatures,
where no phonon absorption processes are possible, the width of the lowest
ZPL is determined by the background lifetime. Outside of the anticrossing
region (left figure) there are no phonon-assisted transitions from the upper
state, either, because of the lack of final states in the phonon-coupled
range. Hence, also the width of the ZPL of the upper line is limited by the
background lifetime. At the anticrossing (right figure), on the other hand,
the upper line exhibits a pronounced additional broadening, because here
phonon-mediated relaxation to the lower branch is more possible. The
sidebands are related to the phonon-assisted absorption of light. For zero
temperatures, only phonon emission can accompany a photon absorption, so that
the sideband appears only at the high energy side of the ZPL. With increasing
temperature, a sideband appears also on the low-energy side and grows until
the sidebands become symmetric at sufficiently high temperature. Slight
modulations of the sidebands again reflect the resonances resulting from the
interplay between phonon wavelength and QD separation.

\subsection{Spin-Orbit Coupling} \label{Res:SOC}
\begin{figure}
    \includegraphics[width=1.0\columnwidth]{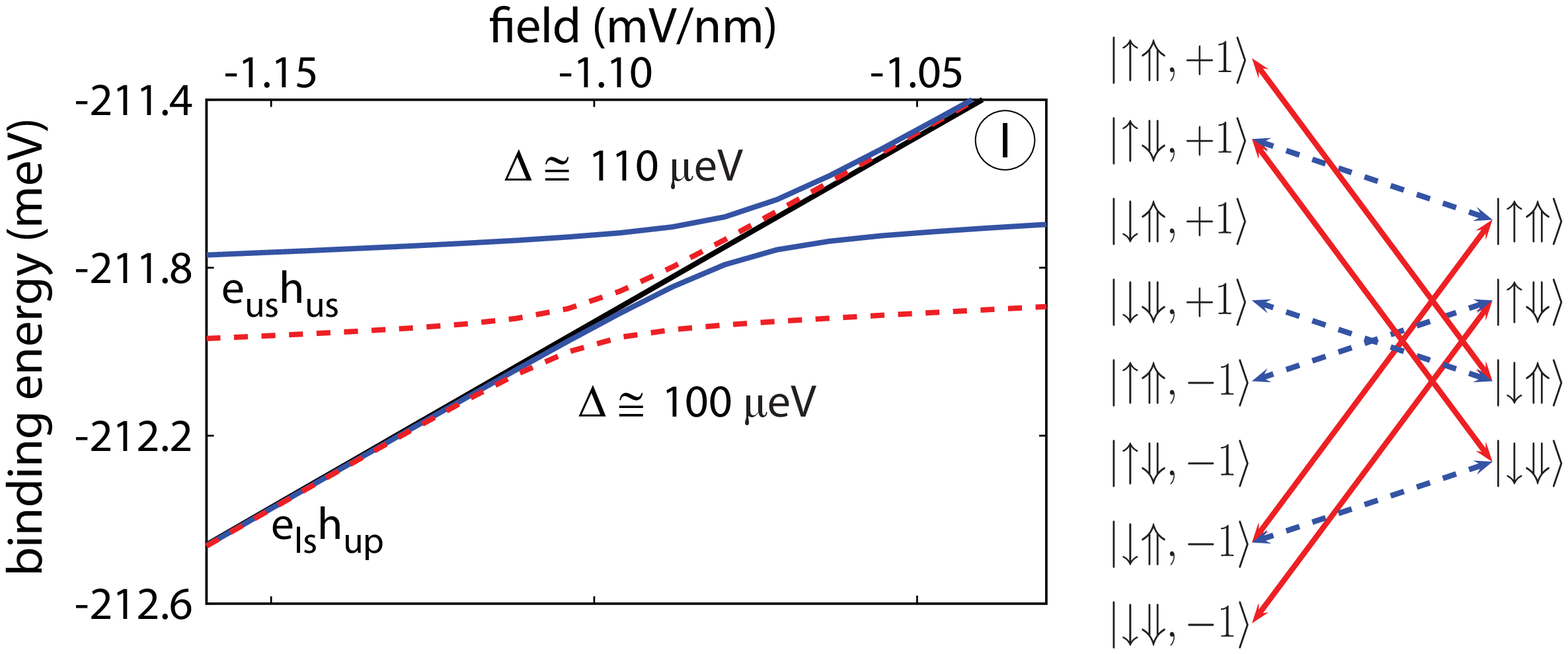}
    \caption{(Color online) Left: Binding energies of the exciton states \X{\us}{\us} and \X{\ls}{\up} as a function of the applied field, including different
    spin configurations and spin-orbit interaction. Right: Scheme of the coupling of the states at the anticrossing due to SOC of electron (red, solid lines) and hole (blue, dashed lines).}
    \label{SOC1}
\end{figure}
For the previous discussions we assumed strict conservation of angular
momentum as a consequence of the cylindrical symmetry of the structure.
Therefore excitons with different values of the angular momentum did not
couple. This conservation is broken by spin-orbit coupling (SOC), although
its effects are typically small. For the crossing of the states \X{\us}{\us}
and \X{\ls}{\up} around $-1$~mV/nm, \Tag{I} in Fig.~\ref{Absorp_LS}, we will
now include SOC in the calculations as described in section \ref{Theo:SOC}.
Figure~\ref{SOC1} (left) shows the corresponding exciton energies as a
function of the applied field. The
direct exciton splits up in two bright ($\ket{\uparrow\Downarrow}$ and
$\ket{\downarrow\Uparrow}$) and two dark ($\ket{\uparrow\Uparrow}$ and
$\ket{\downarrow\Downarrow}$) spin configurations, where the latter two are
shifted downward by $0.2~$meV due to the exchange interaction. For the
indirect exciton we have to consider additionally two different hole
orbital-states with angular momentum of $\pm 1$, adding up to eight states
(Fig.~\ref{SOC1}, right). The indirect-exciton states with
$\ket{\downarrow\Uparrow, +1}$ and $\ket{\uparrow\Downarrow, -1}$ are not
coupled by SOC to the direct-exciton states and remain unaffected (black
line). The remaining states can be grouped into three subsets of coupled
states: $\ket{\uparrow\Uparrow, +1}, \ket{\downarrow\Uparrow},
\ket{\downarrow\Downarrow, +1}$ and $\ket{\uparrow\Uparrow, -1},
\ket{\uparrow\Downarrow}, \ket{\downarrow\Downarrow, -1}$, where the
$s$-state is coupled to the $p$-states by hole or electron SOC (note that the
$p$-state has contributions of the single-particle $p$-state of hole and
electron because of Coulomb interaction). The coupling of these states gives
rise to the upper anticrossing (blue, solid lines). In the last subspace with
$\ket{\uparrow\Uparrow}, \ket{\downarrow\Uparrow,-1},
\ket{\uparrow\Downarrow,+1}$ and $\ket{\downarrow\Downarrow}$, the dark,
direct states are indirectly coupled to each other. This results in the lower
anticrossing (red, dashed lines) with an energy separation of about 100~$\mu$eV. The
effect of SOC vanishes quickly out of resonance. Although the mixing of the
states supports phonon-mediated transitions, the effect is very small, as the
splitting remains in an energy range with a very low spectral density
$J_{nm}(\omega_{nm})$.

\subsection{Confinement potentials with broken cylindrical symmetry} \label{Res:Broken}

\begin{figure}
    \includegraphics[width=1.0\columnwidth]{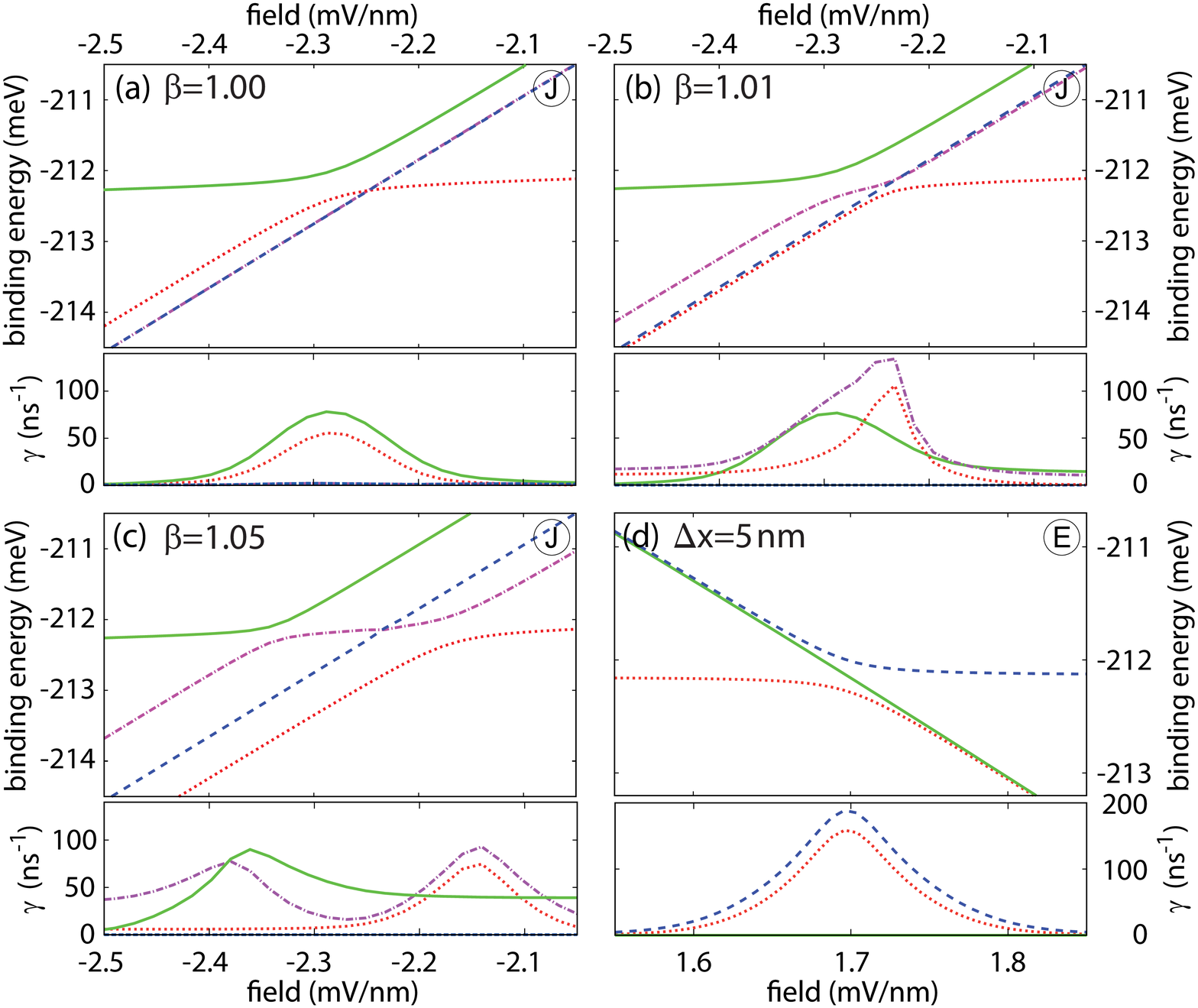}
		\caption{(Color online) Exciton binding energies (top) and phonon-mediated transition
		rates (DP+PE, at 10~K) (bottom), as a function of the applied field for cases with broken
		cylindrical symmetry. (a)-(c): Elliptical dots at \Tag{J}. (d):
		QDs displaced by $5~$nm at \Tag{E}. Transition rates shown in the bottom panels refer to the states represented by the same line style.}
		\label{broken}
\end{figure}
So far, we have considered a structure with cylindrical symmetry, such that
-- in the absence of SOC -- angular momentum has been a good quantum number.
In a real structure the cylindrical symmetry of the confining potentials can
be broken, e.g., if one or both QDs exhibit elliptical elongations in a
specific direction, or in the case of an imperfect alignment, i.e., a lateral
displacement of the center of the two QDs. Such structures with a broken
cylindrical symmetry are described by an additional contribution
$H_{\mathrm{bs}}^{\mathrm{e/h}}$ to the confinement potentials of electrons
and holes. Here we will analyze the role of these types of symmetry breaking
for the coupling between different direct and indirect excitons. We will
assume that the deviations from cylindrical symmetry are not too strong such
that they can be treated in a perturbative way.

In the case of an elliptical deformation of the confinement potentials
degeneracies are removed, e.g., the p-shell splits into $p_x$ and $p_y$.
Between these states phonon-assisted transitions can become rather efficient.
However, a two-fold rotational symmetry axis is still present and no coupling
between $p$- and $s$-state emerges. On the level of the present envelope
function formalism, this selection rule can only be broken if the dot
symmetry is reduced even further, which seems rather unusual.

For a dot with a harmonic in-plane confinement, ellipticity can be described
by a difference in the potentials of $x$- and $y$-direction, i.e., $\omega_x
\neq \omega_y$. Assuming the same deviation from cylindrical symmetry of both
dots, this ellipticity is described in polar coordinates ($\rho, \phi$) by
the Hamiltonian
\begin{equation}
H_{\mathrm{bs}}^{\mathrm{e/h}}=H_{\mathrm{ellip.}}^{\mathrm{e/h}} = \frac{1}{2}
m_{\mathrm{e/h}} \bar{\omega}_{\mathrm{e/h}}^2 \rho^2
\frac{\beta^4-1}{\beta^4+1} \cos(2\phi) , \label{ellip}
\end{equation}
where $\bar{\omega}_{\mathrm{e/h}}^2
=[(\omega^{\mathrm{e/h}}_{\textrm{u}})^2+(\omega^{\mathrm{e/h}}_{\textrm{l}})^2]
/2$ is an average of the confinement frequencies of the upper and lower dot,
and $\beta = \sqrt{\omega^{\mathrm{e/h}}_x/\omega^{\mathrm{e/h}}_y}$
describes the ratio of the $x$- and $y$-axes of the ellipse. We will use the
same form as an effective additional potential and extract the confinement
frequencies $\omega^{\mathrm{e/h}}_{\textrm{u/l}}$ of upper and lower dot
from the splitting between $s$- and $p$-shell states of electrons and holes
in the upper and lower dot, respectively. The asymmetry parameter $\beta$ is
taken to be the same for electrons and holes. Equation~\fref{ellip} indicates
that the perturbation vanishes between all states $i,j$ with $M_i - M_j \neq
\pm 2$.

In the case of a lateral mismatch of the dots, the strict separation of
states with different in-plane parity is broken. For a harmonic confinement
potential a relative displacement by $\Delta x$ of the upper dot with respect
to the lower dot can be introduced by the perturbation
\be
H_{\mathrm{bs}}^{\mathrm{e/h}} = H_{\text{displ.}}^{\mathrm{e/h}}
= m_{\mathrm{e/h}} (\omega_{\textrm{u}}^{\mathrm{e/h}})^2 \left( \frac{\Delta x^2(z)}{2} -\Delta x(z)\,\rho \,\cos(\phi)\right)\label{displ}
\ee
with
\be
\Delta x(z)= \frac{\Delta x}{2} \left[ \tanh \left(\frac{z-z_0}{a}\right) + 1\right], \nonumber
\ee
where the $\tanh$-function ensures a smooth transition of the potential
between the dots, $z_0$ is the $z$-coordinate in the middle between both
dots, and $a$ is a parameter that controls the smoothing, which we take to be
1~nm. Evidently, this perturbation gives rise to a coupling between states
with $M_i - M_j = \pm 1$.

For both types of symmetry-breaking potentials we calculate the matrix
elements in the truncated basis of the single particle states, which leads to
an additional contribution
\begin{equation}
\sum_{ij} \left( H_{\mathrm{bs}}^{\mathrm{e}} \right)_{ij} c\dag_i c\ndag_j +
\sum_{ij} \left( H_{\mathrm{bs}}^{\mathrm{h}} \right)_{ij} d\dag_i d\ndag_j
\end{equation}
to the interacting particle Hamiltonian in Eq.~\eqref{EQN:Coulomb}. From this
extended Hamiltonian the modified exciton energies and wave functions are
then obtained by numerical diagonalization within the basis of electron-hole
pair states.

Figures~\ref{broken}(a)-(c) show the exciton energy of the shells
\X{\us}{\us} and \X{\ls}{\ud} as a function of the applied field for
different elongations around \Tag{J} in Fig.~\ref{Absorp_LS}. For a symmetric
dot, i.e., $\beta=1.00$ [Fig.~\ref{broken}(a)], the degenerate $d$-states
with $M=\pm 2$ (blue, dashed line) are uncoupled from the horizontal
$s$-state and the other $d$-state with $M=0$, which run through an avoided
crossing (red, dotted and green, solid lines). In the presence of an
ellipticity ($\beta \ne 1$) the states with $M=\pm 2$ mix leading to the
symmetry-adapted states with angular dependencies $\sim \cos (2\phi)$ and
$\sim \sin (2\phi)$. By means of the Hamiltonian in Eq.~\eqref{ellip} the
former state now is coupled to $s$-states as well as to $d$-states with $M=0$
while the latter state remains uncoupled. This is indeed seen in
Figs.~\ref{broken}(b),(c) where we observe the lifting of the degeneracy of
the two $d$-states and the appearance of a new anticrossing while the blue,
dashed line, corresponding to the state with the sine-like angular
dependence, still exhibits a crossing behavior. For a ratio of the dot axes
of $\beta=1.01$, the degeneracy of the states is only slightly lifted and a
small additional anticrossing emerges. The splitting between the $d$-states,
as well as the anticrossing, becomes larger for a stronger ellipticity
[$\beta=1.05$ in Fig.~\ref{broken}(c)], while the other anticrossing shrinks.

The lower parts of Figs.~\ref{broken}(a)-(c) show the corresponding
phonon-mediated transition rates. The rates are increased at an anticrossing,
with the higher state having a larger transition rate because of the
relaxation to the lower state. For the coupled $d$-states in
Fig.~\ref{broken}(b) and (c) phonon-assisted transitions occur also out of
the anticrossing region. This is caused by intrashell transitions within the
$d$-shell, which become noticeable for sufficiently large splitting between
the $d$-shell states. Hence, they only depend on the ellipticity of the dots
and stay at a constant level out of the region of an avoided crossing.
Phonon-mediated transitions of the uncoupled $d$-state remain negligible.

Also in the case of a lateral displacement of one of the QDs with respect to
the other the angular momentum eigenstates mix to states with a sine- or
cosine-like angular dependence. This is seen in Fig.~\ref{broken}(d), where
the exciton energies of \X{\us}{\us} and \X{\us}{\lp} are depicted in the
region around the former crossing between $s$- and $p$-states (\Tag{E} in
Fig.~\ref{Absorp_LS}) for the case of a displacement of the upper dot by
$\Delta x = 5$~nm with respect to the lower dot. An anticrossing with a
splitting $\Delta E \approx 300$~$\mu$eV appears due to the coupling between
the $p_x$-state [$\sim \cos (\phi)$] with the $s$-state, while the
$p_y$-state [$\sim \sin (\phi)$] is uncoupled. Here, the degeneracy of the
$p$-shell is only lifted in an area around the avoided crossing, because out
of this region the states are mainly localized in one of the dots and
therefore experience only a very weak deviation from cylindrical symmetry.

The lower part of Fig.~\ref{broken}(d) depicts the transition rates of the
exciton states. The rates of the coupled exciton states are rather high,
since the energy separation at the anticrossing is in the range of an
efficient phonon coupling. For the uncoupled state the transition rates
remain at a very low level.

\section{Conclusion}\label{Conc}

We have investigated excitons in QDMs, in particular their absorption spectra
and the influence of the phonon environment on the spectra. Three different
kinds of anticrossings, which are caused by tunnel coupling, Coulomb
interaction and spin-orbit coupling, were discussed, as well as the influence
of a broken cylindrical symmetry induced, e.g., by an elliptical elongation
of the dots or a lateral displacement of the centers of the QDs.

We have found that the ordering of the energies of the states is strongly
modified by the Coulomb interaction, compared to the sequence of combined
single-particle energies. The Coulomb interaction causes additional
anticrossings between exciton states leading to an enhanced phonon relaxation
in the regions of such anticrossings, comparable to the situation of a tunnel
coupling of the hole. The shape of the absorption lines is characterized by a
zero phonon line superimposed on a broad phonon background. The latter one
results from the pure daphasing-type part of the exciton-phonon coupling,
while the width of the ZPLs can be enhanced by real phonon transitions to
other exciton states, which occurs mainly in the region of an anticrossing.
The resonance between the phonon wavelength and the separation of the QDs
gives rise to characteristic modulations in the phonon scattering rates and
the phonon background.

Although the cylindrical symmetry of the confinement potential supports
conservation of angular momentum, spin-orbit coupling removes this
conservation and leads to more complex anticrossings at an exciton resonance.
The energy splitting introduced by this coupling, however, are typically
rather small such that no efficient phonon scattering between the states
occurs.

In the case of a perturbation of the cylindrical symmetry of the confinement
potential, we have found that an elliptical elongation leads to additional
avoided crossings near resonances between states with angular momenta
differing by two and thus to energy splittings of otherwise degenerate
states. In contrast, a displacement of the center of the dots provides an
additional mixing of otherwise uncoupled states for any difference of the
angular momenta.

\subsection*{Acknowledgements}
This work was supported in part by a Research Group Linkage Project of the
Alexander von Humboldt Foundation and by the TEAM programme of the Foundation
for Polish Science, co-financed by the European Regional Development Fund.
Fruitful discussions with Jonathan Finley, Kai M\"uller, and Krzysztof
Gawarecki are gratefully acknowledged.


\begin{thebibliography}{53}
\bibitem [{\citenamefont {note}(2013)}]{note2013}%
  \BibitemOpen
  \bibfield  {author} {\bibinfo {author} {\bibfnamefont
  { A spatially dependent strain
  distribution is not taken into account explicitly in the calculations since this is expected
  to be of minor importance due to the rather low indium content.
  However, the values for the electron and hole masses are adjusted to reproduce the
  experimental results from Ref.~32 and therefore the overall
  effect of strain is included}} }\BibitemShut {NoStop}%
\end{thebibliography}
\end{document}